\documentclass{aastex631}
\usepackage{amsmath}
\usepackage{bm}
\usepackage{xcolor}
\usepackage{soul}

\received{September 22, 2021}
\revised{January 17, 2023 and April 12, 2023}
\accepted{April 15, 2023}
\submitjournal{\aj}
\shorttitle{Machine learning Kuiper Belt mean plane}
\shortauthors{Matheson \& Malhotra}
\usepackage{ragged2e}
\begin{document}

\title{A Measurement of the Kuiper Belt's Mean Plane From Objects Classified By Machine Learning}

\author[0000-0003-0940-7176]{Ian Matheson}
\affiliation{University of Arizona Department of Aerospace \& Mechanical Engineering \\
1130 N. Mountain Ave. \\
P.O. Box 210119 \\
Tucson, AZ 85721, USA \\
ianmatheson@arizona.edu}

\author[0000-0002-1226-3305]{Renu Malhotra}
\affiliation{University of Arizona Lunar \& Planetary Laboratory \\
1629 E. University Blvd. \\
P.O. Box 210092 \\
Tucson, AZ 85721, USA \\
malhotra@arizona.edu}

\begin{abstract}

Mean plane measurements of the Kuiper Belt from observational data are of interest for their potential to test dynamical models of the solar system. Recent measurements have yielded inconsistent results. Here we report a measurement of the Kuiper Belt's mean plane with a sample size more than twice as large as in previous measurements. The sample of interest is the non-resonant Kuiper belt objects, which we identify by using machine learning on the observed Kuiper Belt population whose orbits are well-determined. We estimate the measurement error with a Monte Carlo procedure. We find that the overall mean plane of the non-resonant Kuiper Belt (semimajor axis range 35-150 au) and also that of the classical Kuiper Belt (semimajor axis range 42-48 au) are both close to (within $\sim0.7^\circ$) but distinguishable from the invariable plane of the solar system to greater than 99.7\%
confidence. When binning the sample into smaller semimajor axis bins, we find the measured mean plane mostly consistent with both the invariable plane and the theoretically expected Laplace surface forced by the known planets. Statistically significant discrepancies are found only in the semimajor axis ranges 40.3-42 au and 45-50 au; these ranges are in proximity to the $\nu_8$ secular resonance and Neptune's 2:1 mean motion resonance where the theory for the Laplace surface is likely to be inaccurate. These results do not support a previously reported anomalous warp at semimajor axes above 50 au.

\end{abstract}


\section{Introduction}

\cite{cc08} posed the question:``If we could map, at fixed time, the instantaneous locations in three-dimensional space of all Kuiper Belt objects [KBOs], on what two-dimensional surface would the density of KBOs be greatest?'' The authors demonstrated that this surface, also known as the Laplace surface, is given by the Laplace-Lagrange linear secular theory \citep{md99}. This theory is based on the time-variable forcing arising from the planets' secular variations; consequently, the local normal on the Laplace surface varies only slowly with time; secular timescales for KBOs are much longer than $\sim 10^4$ yr.
The Laplace surface for particles within the Kuiper Belt is not a flat plane because it has warps owing to secular resonances in certain localized regions of semimajor axes within the belt where the rate of orbit pole precession coincides with one of the inclination secular mode frequencies of the planets; at large semimajor axes the Laplace surface converges to the  solar system's invariable plane. The invariable plane is the flat plane normal to the total orbital angular momentum of the solar system; this plane has an inclination of $1.58^\circ$ and a longitude of ascending node of $107.58^\circ$ with respect to the J2000 ecliptic-equinox frame \citep{ss12}.

As a foil to the Laplace surface, previous studies have also considered the solar system's invariable plane as a candidate for the mean plane of the Kuiper Belt. The mean plane measured from observational data of KBOs is of interest for its potential to test dynamical models of the solar system and to reveal unmodeled perturbations when compared with theoretical predictions.

Previous measurements of the Kuiper Belt mean plane have produced inconsistent results.  \cite{bp04} reported a mean plane of $(i_0,\Omega_0)=(1.86^\circ,81.6^\circ)$, where $i_0$ and $\Omega_0$ are respectively the inclination and longitude of ascending node of the plane in J2000 coordinates.  With a measurement error of $0.37^\circ$ in the pole position of the mean plane, this is consistent with the Laplace surface at semimajor axis $a=44$ au (the median semimajor axis of their sample) but inconsistent (at more than 99.7\% confidence) with the invariable plane of the solar system as well as the orbital planes of Neptune and Jupiter.  \cite{e05} used five separate methods to measure the mean plane of the Kuiper Belt; they reported $i_0$ in the range $1.65^\circ - 2.49^\circ$ and $\Omega_0$ in the range $97.4^\circ - 104.0^\circ$, with a preferred value of $(i_0,\Omega_0)=(1.51^\circ\pm0.26^\circ,100.0^\circ\pm8.8^\circ)$.  They rejected the  Laplace surface at 99.7\% confidence as the Kuiper Belt mean plane, but did not reject the invariable plane.     \cite{cc08} measured the mean plane for two small samples of KBOs near $a=38$ au and $a=43$ au and concluded that they could not reject either the Laplace surface or the invariable plane for either sample.  After a gap of more than a decade, the next measurement of the Kuiper Belt's mean plane was reported in \citet{vm17} (hereafter \textsc{vm17}) when the observed sample of Kuiper Belt objects with well-determined orbits ($<5$\% semimajor axis uncertainty, observed over three or more oppositions) had grown to 931. These authors carefully identified resonant KBOs and discarded them from the mean plane measurement sample (because the assumptions of Laplace theory are violated for particles in mean motion resonances).  \textsc{vm17} reported an overall mean plane of $(i_0,\Omega_0)=(1.8^\circ,77^\circ)$ for the classical Kuiper Belt, i.e., the sample of non-resonant KBOs in the semimajor axis range 42--48 au.  For smaller semimajor axis bins, their mean plane measurements in the semimajor axis range of 35 au to 45 au were consistent with the Laplace surface, notably including the detection of the theoretically predicted warp near $a=40$~au; save for that warp, the mean plane in this range was also consistent with the invariable plane. However, beyond $a=50$~au, \textsc{vm17} measured mean planes that were strongly warped away from the predicted Laplace surface and inconsistent with both the Laplace surface and the invariable plane at the  97\%--99\% confidence level.  The most recent measurement of the Kuiper Belt's mean plane, by \cite{ossos14}, was based on a sample of KBOs discovered in two specific observational surveys with well-characterized selection biases. These authors rejected the invariable plane and did not reject the Laplace surface for semimajor axes below $44.4$~au; at larger semimajor axes both planes were accepted.  The significant warp of the mean plane of the distant Kuiper Belt, beyond $a\approx50$, measured by \textsc{vm17} but not detected by \cite{ossos14} is an open puzzle in the literature.

\begin{figure}[htb!]
	\centering
	\includegraphics[width=5.5in]{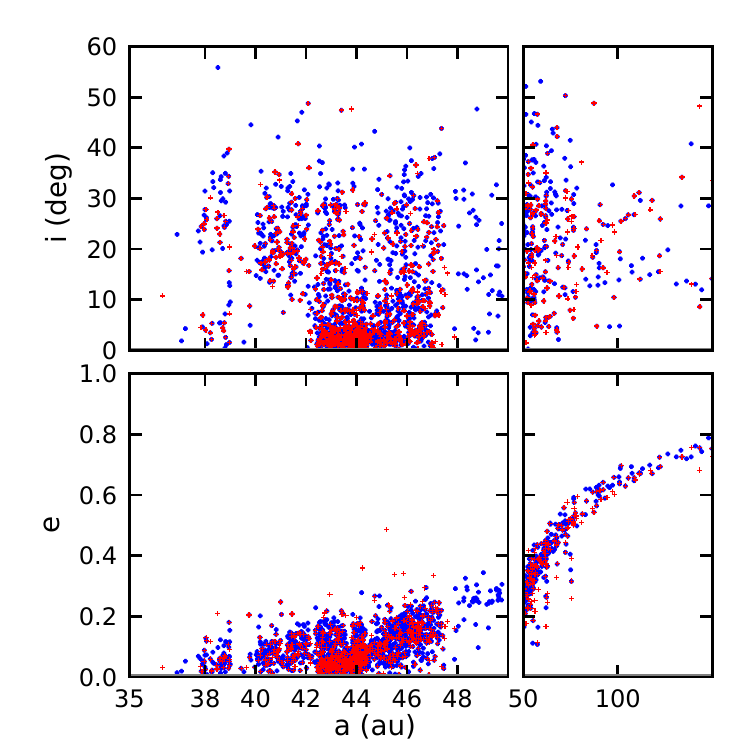}
	\caption{Semimajor axis, inclination, and eccentricity of the 931 non-Resonant KBOs with well-determined orbits on $35<a<150$ au from \textsc{vm17} (red +), and the sample of 1812 non-Resonant KBOs with well-determined orbits on $35<a<150$ au, $q>30.34$ au, in this study (blue $\circ$). Orbital elements for the 1812 non-Resonant KBOs are from \citet{jpl_sbdb_api} as explained in Section \ref{sec:sample_selection}.}
	\label{fig:ai_plot}
\end{figure}

Since \textsc{vm17} published their work, the sample of known non-resonant KBOs with well-determined orbits, semimajor axes between 35--150 au, and perihelia above the semimajor axis of Neptune has nearly doubled, growing from 931 to 1812. Figure \ref{fig:ai_plot} shows a scatter plot of the semimajor axes and ecliptic inclinations of these samples. In the present work, we revisit the measurement of the Kuiper Belt's mean plane with this larger sample with the goal to test the reproducibility of \textsc{vm17}'s results.  In Section 2,  we briefly describe the theoretical background for the Laplace surface. In the next two sections, we describe the method for measuring the mean plane from observations with generally unknown biases (Section 3), and a Monte Carlo method for estimating the uncertainty of this measurement (Section 4). In Section 5, we describe the machine-learning tool for identifying the sample of non-resonant KBOs. Our results are presented and discussed in Section 6.

\section{Laplace surface}

Though their measurements of the mean plane differ, researchers including \textsc{vm17}, \cite{cc08}, and \cite{bp04} all start with the premise that for KBOs with perihelia outside the orbit of Neptune, the theoretically expected mean plane is adequately described by Laplace theory, i.e., the linear secular perturbation theory as set forth in \cite{md99}.  In this approximation, the planetary perturbations on a test particle are orbit-averaged and the perturbing potential (called the ``disturbing function") is truncated to the first order in the planetary masses and the second order in eccentricities and inclinations.  It then follows from Lagrange's equations for the variation of the orbital elements that the test particle's semimajor axis is constant, while its orbital eccentricity and the orientation of its orbit vary quasi-periodically with time over secular timescales.  Our interest here is in the orbit plane, which is described by two angular elements, the inclination, $i$, and the longitude of ascending node, $\Omega$. The unit vector, $\hat{\bm{h}}$, normal to the orbit plane can be expressed in terms of these elements (Eq.~\ref{e:hat_h}). The Laplace theory expositions in the literature use the more convenient two-component vector, the so-called ``inclination vector" defined as

\begin{equation}
(q,p) = \sin i\, (\cos\Omega,\sin\Omega).
\end{equation}

Laplace theory yields the solution of the linear secular equations for the test particle in which $(q,p)$ can be written as a sum of two parts,

\begin{equation}
(q,p) = (q_0,p_0) + (q_1,p_1) .
\end{equation}

The first part, $(q_0,p_0)=\sin i_0\,(\cos\Omega_0,\sin\Omega_0)$, is called the ``forced inclination vector" and defines the Laplace surface; it is determined by the masses and semimajor axes of the planets and their instantaneous orbital planes, as well as the semimajor axis of the test particle. The second part, $(q_1,p_1)=\sin i_1\,(\cos\Omega_1,\sin\Omega_1)$, called the ``free inclination vector," is the remaining part of the total inclination vector; it is determined by initial conditions. The Laplace surface changes over secular times, and its time variation is given by a superposition of the secular modes of the inclination vectors of the planets. The free inclination vector precesses around the Laplace surface at a constant angular rate: the free inclination $i_1$ remains constant and the free longitude of node $\Omega_1$ circulates.  As the orbit plane of a KBO precesses around its local Laplace surface, it follows that the mean plane of a large population of test particles in a small semimajor axis range (with  dispersed orbit planes) is described by the local Laplace surface, as shown in simulations by \cite{md99} and \cite{cc08}.

For later reference, we note that the unit vector, $\hat{\bm{n}}_0$, normal to the local Laplace surface is related to the forced inclination vector as follows,

\begin{equation}
	\hat{\bm{n}}_0=\bigg(p_0,-q_0,\sqrt{1-p_0^2-q_0^2}\bigg).
\end{equation}

We briefly make a note about heliocentric versus barycentric orbital elements.  The Laplace surface on 2023 February 20 for semi-major axes outside Neptune's orbit, shown in Figure \ref{fig:iOmega}, is computed according to the theory described in \cite{md99}, and using data for the J2000 ecliptic/equinox barycentric planetary elements retrieved from JPL Horizons and planetary masses from  \cite{standish1995report}.  Laplace theory as set forth in \cite{md99} and other sources describes the Laplace surface in heliocentric coordinates, rather than barycentric coordinates.  The portion of the disturbing function used to develop the theory depends on eccentricity, inclination, longitude of node, longitude of pericenter, and semimajor axis.  Only semimajor axis is affected by the heliocentric-to-barycentric conversion. The heliocentric semimajor axis has short-period oscillations caused by the motions of planets interior to the Kuiper Belt; these average to zero over secular timescales. Barycentric elements, which lack those semimajor axis oscillations, are therefore more convenient and no less accurate than heliocentric coordinates for high-$a$ Laplace surface calculations. As a practical matter, the Laplace surface for orbits outside Neptune's is nearly identical in heliocentric and barycentric coordinates, except in close proximity to the secular resonance locations near (barycentric) semimajor axis values of 35 au and 40.3 au.

\section{Mean Plane Measurement} \label{sec:mean_plane_measurement}

At first sight, it would seem reasonable to simply average the total inclination vectors of a group of KBOs to find their mean plane. Stated more carefully, one could average the unit vectors normal to the orbit planes of the KBOs to find the unit vector normal to their mean plane. However, without a complete or fair sample, the result of such a calculation will reflect the biases of the observational surveys that produced the sample \citep{bp04,e05,vm17}. Observational surveys not only have brightness limits, but surveys of the outer solar system for KBOs also have limited and non-uniform coverage in ecliptic latitude and ecliptic longitude. The resulting KBO samples will yield average planes biased toward the observed regions, whatever the true population mean plane may be (\textsc{vm17}).

To mitigate the observational biases in catalogs of KBOs, \cite{bp04} proposed that the unbiased mean plane of the Kuiper Belt could be identified with the plane of symmetry of the vectors, ${\bm{v}}_t$, of KBOs' space velocities projected on the heliocentric celestial sphere; ${\bm{v}}_t$ can be called the sky-plane velocity vector. The instantaneous unit vector, $\hat{\bm{v}}_t$, along the direction of a KBO's sky-plane velocity vector can be computed from the KBO's observational data pertaining to its orbital plane and its position in the sky at some epoch, as follows. Denoting with $\hat{\bm{h}}$ the unit vector normal to the orbit plane, and with $\hat{\bm{r}}$ the unit vector along the radial direction to the KBO's heliocentric position, then

\begin{equation}
	\hat{\bm{v}}_t=\hat{\bm{h}}\times\hat{\bm{r}}.
\label{e:hat_v}\end{equation}

The unit vector, $\hat{\bm{h}}$ is found from a KBO's inclination and longitude of ascending node, $i$ and $\Omega$, and the unit vector $\hat{\bm{r}}$ is found from its J2000 ecliptic latitude  and ecliptic longitude, $\beta$ and $\lambda$, follows,

\begin{eqnarray}
	\hat{\bm{h}}
	&=&\big(\sin i\sin\Omega,-\sin i\cos\Omega,\cos i\big),
\label{e:hat_h} \\
	\hat{\bm{r}}
	&=&\big(\cos\beta\cos\lambda,\cos\beta\sin\lambda,\sin\beta\big).
\label{e:hat_r}
\end{eqnarray}

\textsc{vm17} described a simpler implementation of the method of \cite{bp04} for computing the mean plane of a KBO sample from the sky-plane velocity vectors.  The sky-plane velocity of a KBO orbiting exactly on the mean plane will always be normal to the unit vector $\hat{\bm{n}}$ defining the normal to the mean plane, so that

\begin{equation}
	\hat{\bm{n}}\cdot\hat{\bm{v}}_t=0.
\end{equation}

The mean plane can then be computed as the plane whose orbit normal is \textit{most} normal to the \textit{most} KBO sky-plane  velocities, i.e., the plane that minimizes

\begin{equation}
	S=\sum_{i=1}^{N} |\hat{\bm{n}}\cdot\hat{\bm{v}}_{t,i}|.
\end{equation}

The function $S$ may be minimized on a grid search in the $(q,p)$ parameter space for a computationally simple, though grid-dependent, method of finding the mean plane.  That is, we search on a 700~x~700 grid with $q_\textrm{min}=p_\textrm{min}=-0.35$, $q_\textrm{max}=p_\textrm{max}=+0.35$. The precision of the mean plane identification is limited by the grid spacing. With our choice of grid, the precision of the pole position of the computed mean plane is $\lesssim0.1$~degrees. This is small compared to the confidence intervals that are computed later. The mean plane $(q_0,p_0)$ is trivially converted to an inclination and longitude of node $(i_0,\Omega_0)$.

\cite{bp04} and \textsc{vm17} showed, with simulations of biased synthetic samples, that this method of estimating the midplane by using sky-plane velocity vectors substantially mitigates the systematic error and recovers the true mean pole more reliably than simply averaging the orbit pole unit vectors. For example, when this method is applied to a synthetic sample confined to only a small patch of the sky, it recovers the true mean pole whereas the average of the unit pole vectors is systematically offset from the true mean pole (see, e.g., Fig.~8 in \textsc{vm17}). A present disadvantage of this method is that the confidence limits of the measured mean plane are not straightforward to compute (discussed in the next section). The performance of this method and its limitations and reliability with different survey designs merit further exploration.

\section{Mean Plane Measurement Uncertainty} \label{sec:uncertainty}

The method of measuring the mean plane as the plane of symmetry of the sky-plane motion vectors does not render itself to a straightforward method for computing the measurement uncertainty. We follow the method of \cite{bp04} and \textsc{vm17}, who adopted a heuristic approach which assumes a statistical model of the intrinsic inclination distribution rather than fitting for it, and constructs Monte Carlo simulations to generate synthetic samples which approximately account for the biases in the observed sample.
In brief, this approach is as follows.  For a given observational sample of $N_K$ non-resonant KBOs, we generate a new list of $N_K$ simulated objects. The simulated objects are assigned orbital elements drawn randomly from a statistical model of their intrinsic distribution. Each random draw is accepted or rejected by subjecting it to a comparison with the properties of the observed sample; this approximately accounts for the selection biases in the observational data. With a simulated sample in hand, we compute its mean plane, and repeat $N_R$ times with additional simulated samples. In this way, we generate the statistics of the mean plane. We use the distribution of these $N_R$ Monte-Carlo-sampled mean planes to estimate the 68.2\%, 95.4\%, and 99.7\% confidence error bars on the measured mean plane. A more detailed description of these Monte Carlo simulations is given below.

In each of the $N_R$ simulated samples, the $i$th simulated object is found by first randomly selecting the $j$th real object from the observational KBO sample, where $j$ is an integer randomly selected from the discrete uniform distribution on $[1,N_K]$. (How the objects are ordered, $1-N_K$, does not matter.)  We then choose its semimajor axis $a_i$ randomly from the continuous uniform distribution on $[0.99a_j,1.01a_j)$, and we choose its eccentricity $e_i$ randomly from the continuous uniform distribution on $[0.95e_j,1.05e_j)$, so that the model population has approximately the same semimajor axis and eccentricity distribution as the observed population.

The most important assumption is a model for the intrinsic distribution of free inclinations. Our current best understanding of the intrinsic inclination distribution in the Kuiper Belt is found in estimates obtained from a few well-characterized observational campaigns, such as the Canada-France Ecliptic Plane Survey (CFEPS, \cite{Petit:2011,Petit:2017}) and the Outer Solar Systems Origins Survey (OSSOS, \cite{Bannister:2018}). The data from these surveys is more effectively debiased with the use of survey simulators. It is then fit to physically-motivated models for the intrinsic inclination distribution function. These models use the Rayleigh distribution function, or a combination of two Rayleigh distributions, one with a narrow width and one with a wider width, to accommodate models for different dynamical classes of KBOs \citep[e.g.][]{brown_2001,Petit:2011}. \textsc{vm17} adopted these models for the intrinsic distribution of free inclinations; we refer the reader to this paper for a detailed description. \cite{ossos14} obtained debiased estimates with slightly updated data and reported very similar results for the intrinsic inclination distribution models.

For the purposes of our goal to test and update the results of \textsc{vm17} with the larger current observational sample of catalogued KBOs, we follow the choices made in that work for the functional form and parameters of the model distribution of free inclinations. Thus, as in \textsc{vm17}, for semimajor axis bins above 50 au we draw free inclinations from a Rayleigh distribution with scale parameter $\sigma=\sin 18^\circ$; below 50 au, the free inclinations are drawn from a mixture of two Rayleigh distributions with scale parameters  $\sigma_1=\sin 3^\circ$, $\sigma_2=\sin 13^\circ$.  If the selected $j$th real object has inclination below $5^\circ$ and a random number drawn from the continuous uniform distribution on [0,1) is less than 0.9, we set $\sigma=\sigma_1$, otherwise $\sigma=\sigma_2$.  \textsc{vm17} used this rule to generate simulated objects approximately evenly split between the low-inclination and high-inclination Rayleigh distributions. The longitude of ascending node of the free inclination vector is drawn randomly from the continuous uniform distribution on $[0,2\pi)$. The total inclination vector of the simulated object is then computed as the sum of this randomly generated free inclination vector and the inclination vector of the mean plane of the $N_K$ objects in the observational sample, as computed in Section \ref{sec:mean_plane_measurement}; that is,

\begin{equation}
	(q_i,p_i)=(q_0,p_0)+(q_{1,i},p_{1,i}),
\end{equation}

where $(q_0,p_0)$ is the inclination vector of the mean plane of the $N_K$ observed objects and $(q_{1,i},p_{1,i})$ is the randomly generated free inclination vector.  The remaining angular orbital elements -- the mean anomaly and the argument of pericenter -- are each drawn randomly and independently from the continuous uniform distribution on $[0,2\pi)$. With all the orbital elements of the simulated object in hand, we then compute its ecliptic latitude $\beta_i$ and ecliptic longitude $\lambda_i$.  Simulated objects are generated until there is exactly one simulated $(\beta_i,\lambda_i)$ pair for each $(\beta,\lambda)$ pair in the observational sample, where, as in \textsc{vm17}, an acceptable match is within one degree in $\beta$ and within five degrees in $\lambda$. Unfortunately, some objects (typically one or two per semimajor axis bin) are difficult to match in this manner. If an object has not been matched after $10^7$ random draws, we accept the following draw and move on to the next object. As explained in \textsc{vm17} and \cite{bp04}, this procedure of generating the synthetic samples by matching their sky locations to those of the real observational sample approximately accounts for the selection biases present in the real observational sample, subject to the key assumption that each patch of sky in ecliptic coordinates (where a real object is found) has been thoroughly observed.  It follows from this assumption that a true population will have the same number of objects in each patch of sky as the observed catalog, and we roughly control for the brightness and magnitude biases of the many surveys that comprise the entire KBO catalog by requiring the semimajor axis and eccentricity of each synthetic object to be close to that of a real object.

Having generated $N_K$ simulated objects, we compute the mean plane of the simulated sample as described in Section \ref{sec:mean_plane_measurement}.  We repeat this procedure $N_R$ times, producing a set of simulated mean planes $(q_{0,r},p_{0,r})$, where $r$ is an integer from 1 to $N_R$. We use the \textsc{R} command \textsc{dataellipse} to compute the 68.2\%, 95.4\%, and 99.7\% covariance ellipses of the set of simulated mean planes. We are now ready to produce uncertainty intervals for the inclination and longitude of ascending node of the mean plane of the observational sample, $i_0$ and $\Omega_0$.  If these were independent quantities defined on infinite domains, we would simply report low and high percentiles from the inclinations and longitudes of ascending node of the $N_R$ synthetic mean planes, but both $i_0$ and $\Omega_0$ have finite angular domains. Instead, we use the 68.2\% (95.4\%, 99.7\%) covariance ellipse from the simulated mean planes $(q_{0,r},p_{0,r})$ and take the maximum and minimum values of $i_0$ and $\Omega_0$ on the ellipse as the 68.2\% (95.4\%, 99.7\%) confidence intervals for the inclination and longitude of node of the mean plane of the observational sample.  If the covariance ellipse surrounds the origin in the $(q,p)$ plane, the confidence interval for $\Omega_0$ spans the entire circle, $[0,2\pi)$, and the confidence interval for $i_0$ has its lower endpoint at zero inclination.  We also report a single number, $\sigma_i$, to describe the uncertainty of the pole position, in degrees, on the celestial sphere, computed as

\begin{equation}
	\sigma_i = \arcsin{\frac{1}{2}(a+b)},
\end{equation}

where $a$ and $b$ are the semimajor and semiminor axes of the 68.2\% covariance ellipse of the simulated mean planes.

For each semimajor axis bin, $N_R=40,000$ synthetic mean planes are computed, which is sufficient to get convergence of the confidence intervals.
Note that our grid search computes $i_0$ and $\Omega_0$ to a precision of $\sim0.1^\circ$ and $\sim1^\circ$, respectively, but our Monte Carlo procedure produces 68.2\% confidence intervals of these parameters that are an order of magnitude larger; the latter is the 1$\sigma$ uncertainty of the measured mean plane.

\section{Sample Selection} \label{sec:sample_selection}

Laplace theory has been developed with the assumption that no two planets in the system are in mean motion resonances with each other, and no test particle is in a mean motion resonance with any planet.  Before the mean plane of any observational population of KBOs can be computed for comparison with the Laplace surface, all resonant KBOs must be identified and removed.

To identify the suitable sample of KBOs for which to calculate the mean plane, we first used the JPL Solar System Dynamics Group's Small Body Database Query to retrieve all asteroid-type (as opposed to comet-type) objects with $e<1$ and heliocentric semimajor axis constrained to $30<a<200$~au.  On 2023 February 20, this returned heliocentric elements for 4149 objects.
We eliminated all objects with fractional semimajor axis uncertainty $\sigma_a/a>5\%$, as well as the eleven non-cometary objects where the semimajor axis uncertainty was unstated. Next, we downloaded the MPCORB.DAT database from the MPC on 2023 February 20 and cross-referenced it against the Small Body Database to eliminate all objects that have been observed for fewer than the three oppositions recommended by \cite{gmv08} and eliminated all objects with cometary designations.  For each of the remaining objects, we retrieved barycentric elements at 2023 February 20 from JPL Horizons using the Python package \textsc{Astroquery} \citep{2019AJ....157...98G}.
We then eliminated any remaining comets according to the criteria in \citet{gmv08}, i.e. objects with a Tisserand parameter of $T_J<3.05$ and a perihelion of $q<7.35$~au. Next, we enforced barycentric semimajor axis limits of $a<150$~au. We introduced a perihelion cutoff of $q>30.34$ au (which is equal to the barycentric aphelion distance of Neptune on 2023 February 20) to eliminate any remaining planet-crossing objects. We then eliminated objects with barycentric $a>150$ au. After the initial stages of sample selection, 2810 objects remained. We classified the remaining objects as Resonant or non-Resonant as described below, and eliminated the Resonant objects from the sample.

The defining property of a resonant object is the libration of its critical resonance angle.
The generally accepted method for identifying resonant KBOs in a sample population is to integrate the orbit of each KBO in an accurate \textit{n}-body integrator, including perturbations from all the planets, recording all the resonant angles of interest, and then examine plots of the time series of the resonant angles by eye to look for persistent librations.  The orbit must be integrated long enough to detect the libration of the longest-period resonant angles. Standard resonance identification methods, most notably that of \citet{gmv08}, integrate for 10 Myr. For the longest-period angles, for angles that librate over large fractions of the circle, or for objects that alternate between libration and circulation, it may be difficult to distinguish resonant objects from non-resonant objects.  Uncertainty in KBO orbital elements can also complicate resonant classification.

The current count of KBOs with well-determined orbits to classify (2810 objects) is rather large for efficient classification by eye. When assigning resonant classification by eye, it is important to maintain mental consistency so that objects at the beginning, middle, and end of the list receive the same scrutiny. Consistency of resonant evaluations across the sample may be improved by repeatedly shuffling the sample and re-classifying the objects, then accepting the most common classification for each object, or by deliberately evaluating each object according to a checklist of features that can be seen by eye. If a checklist is to be used, however, it is faster, more reliable, and more reproducible to automate the process by encoding those features in a classification algorithm (one example is \citet{yu18}). The machine-classified objects may then be re-examined by eye to check for errors, if desired. Because new generations of telescopes are expected to dramatically increase the number of known KBOs in the upcoming decade, it is desirable to have an automated method to classify KBOs.

\cite{sv20} used the criteria of \cite{gmv08} to classify 2305 KBOs from the Minor Planet Center (MPC) database as of 2016 October 20 after fitting new orbits to each object and integrating them for 10 Myr.  They then used the Python machine learning package \textsc{Scikit-Learn} (\cite{scikit-learn}) to develop a gradient boosting classifier for classification of KBOs as either Resonant, Classical, Detached, or Scattering, training said classifier upon the orbits newly classified using the \cite{gmv08} criteria.  Their code integrates a KBO orbit from initial barycentric elements in the \textit{n}-body integrator \textsc{rebound} (\cite{rebound}) for 100 kyr and records 55 features of the orbit for use by the machine learning algorithm.  Full details of the settings used for the machine learning algorithm, and a full explanation and justification of the 55 recorded features, are given in their paper. This machine learning classifier was demonstrated to reproduce with 97\% accuracy the classifications of their testing sample of 542 objects while using orbit integrations of only 1\% the length of the 10 Myr standard. \cite{sv20} posted user-friendly \textsc{python} sample code and training data to GitHub to allow other researchers to use their gradient boosting classifier to classify KBOs by simply providing the barycentric orbital elements as inputs.

We downloaded the \textsc{python} sample code and training data (KBO\_features.csv) from the \cite{sv20} GitHub repository.  We used their gradient boosting classifier without modification, trained it on the same training set they provided, exactly as in the sample code, and used it to classify our sample of 2810 KBOs remaining after the down-selection procedure described above.

To account for orbital uncertainties, we used the JPL Small Body Database API \citep{jpl_sbdb_api} to download a JSON file for each object to be classified. The JSON file contained a nominal heliocentric orbital state, a 6x6 covariance matrix for the heliocentric orbit, and an epoch for the nominal orbit and the covariance matrix. The heliocentric orbital elements and their covariance were given as $e,q,t_p,\Omega,\omega,i$, i.e. eccentricity, perihelion distance in au, time of perihelion passage (Julian date), longitude of the ascending node, argument of perihelion, and inclination, where all angles are in degrees and referenced to the J2000 plane, and the epoch is a Julian date. JSON covariances for two of the 2810 objects were unavailable, so we could only classify 2808 objects.

We generated 301 heliocentric orbital element sets for each object: the nominal orbit, and 300 clones from a Gaussian distribution centred at the nominal orbit, from the given covariance. The mean anomaly for each orbital element set was computed as the mean motion for the semimajor axis, times the elapsed time between the time of perihelion passage and the epoch. We used Horizons to download heliocentric orbital elements for the giant planets at each epoch. For each of 2808 objects, we then built 301 \textsc{rebound} simulations consisting of the outer planets at the appropriate epoch and the orbital element set of the nominal orbit or the clone at the same epoch, treating the planets as massive particles and the Kuiper belt object or the clone as a massless test particle.

Each \textsc{rebound} simulation was then classified as Classical, Scattering, Detached, or Resonant using the unmodified \citet{sv20} gradient boosting classifier. Of the 2808 cloned objects, 1812 had zero clones classified as Resonant, 304 had 1--300 clones classified as Resonant, and 692 had all 301 clones classified as Resonant. To be absolutely sure no resonant objects contaminated our sample, we only accepted the 1812 objects for which no clones were Resonant. Had we selected a different cutoff (50\% Resonant clones), our sample size would have been 150 objects larger. A further classification of the non-Resonant objects as Classical, Detached or Scattering was not needed for our mean plane computations, so we did not further examine the classifications of the clones. Mean plane calculations then proceeded using the nominal orbits of the remaining objects.  The complete set of 1812 non-Resonant KBOs is provided online.  A small sample is shown in Table \ref{table:bin_30_150}.  Non-Resonant KBO counts are given by semimajor axis bin in Table \ref{table:vm17method_results}. This table also contains KBO counts by bin for \textsc{vm17}, for comparison. The online supplementary material for this paper includes a ZIP archive with the JSON files that contain the JPL covariance matrix for each object.

\begin{deluxetable}{lrrrrrrrrr} \label{table:bin_30_150}
	\tablecaption{Sample Table of Non-Resonant KBOs.}
	\tablehead{\colhead{MPC ID} & \colhead{$a$ (au)} & \colhead{$e$} & \colhead{$i$} & \colhead{$\Omega$} & \colhead{$\omega$} & \colhead{M} & \colhead{$r$ (au)} & \colhead{$\beta$} & \colhead{$\lambda$}}
	\startdata
	15760 & 43.9 & 0.07 &  2 & 359 &   3 &  36 & 41.5 &   2 &  43 \\
	15807 & 43.7 & 0.06 &  1 & 177 & 305 &  82 & 43.6 &   0 & 211 \\
	15874 & 83.4 & 0.58 & 24 & 218 & 185 &  10 & 39.1 & -18 &  84 \\
	15883 & 47.0 & 0.21 & 19 & 127 & 301 &  76 & 46.6 &  12 & 166 \\
	16684 & 44.1 & 0.05 &  4 &  26 & 252 & 333 & 42.1 &  -2 & 247 \\
	\enddata
	\tablecomments{Non-Resonant KBOs from 34-150 au, from the JPL and MPC databases as of 2023 February 20, in barycentric elements at the epoch of 2023 February 20. All angles are given in degrees. $a$, $e$, $i$, $\omega$, $M$, and $r$ are provided by JPL Horizons, and $\beta$ and $\lambda$ are calculated from Eq.~\ref{e:hat_r}. The complete table is available online.}
\end{deluxetable}
Note that Table \ref{table:bin_30_150} is given for illustration only. Neither the low precision of the orbital elements in the printed table nor the high precision of the elements in the electronic table represents the true statistical precision of the JPL Horizons orbits. This is because the uncertainties reported in the JSON covariance matrix are not straightforwardly related to the uncertainties of the orbital elements reported in Table \ref{table:bin_30_150}. The JPL covariance matrix is given in a different set of orbital elements that must be transformed to the standard set ($q$ to $a$, $t_p$ to $M$). Table 1 also contains elements ($r$, $\lambda$, $\beta$) that are not linearly related to the standard set. Also, the JPL covariance matrix is given at a different epoch for each object, and Table \ref{table:bin_30_150} reports the elements of each object at a common epoch. It is well known that Gaussian uncertainty regions around a nominal orbit do not stay Gaussian as the nominal orbit is integrated forward or backwards in time, although the deviation from Gaussianity is small for small elapsed times.

\section{Results and Discussion} \label{sec:results}

\begin{figure}[htb]
	\plotone{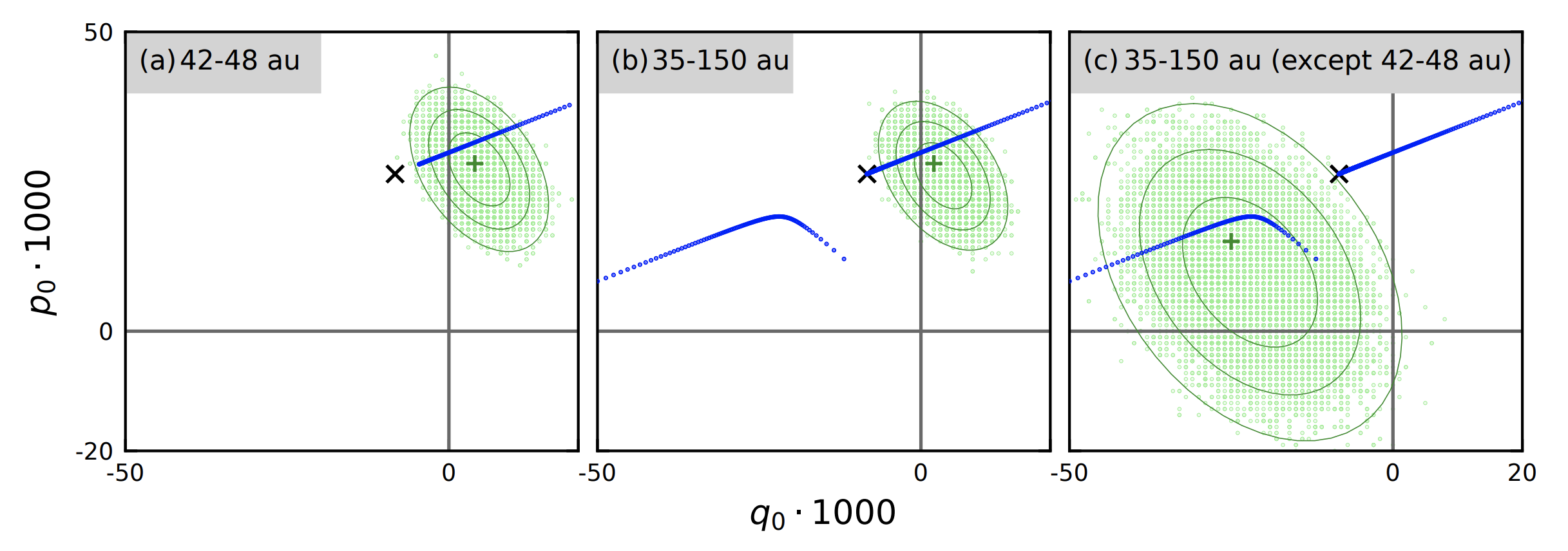}
	\caption{Kuiper Belt mean plane and its confidence ellipses in the $(q,p)$ plane, by semimajor axis bin.  Semimajor axis bins are (a) the classical belt from 42-48 au, (b) the entire belt from 35-150 au, and (c) the entire belt minus the classical region. The best-fit mean plane is the dark green +.  The mean planes of 40,000 Monte-Carlo samples are in light green, and the 68.2\%, 95.4\%, and 99.7\% covariance ellipses for them are in dark green. The J2000 ecliptic/equinox pole is at the origin.  The invariable plane is the black x. The theoretical prediction (from linear secular theory) for the Laplace surface as a function of semimajor axis is plotted in blue. More details are given in the main text.
 }
	\label{fig:qp_2}
\end{figure}

\begin{figure}[htb]
	\plotone{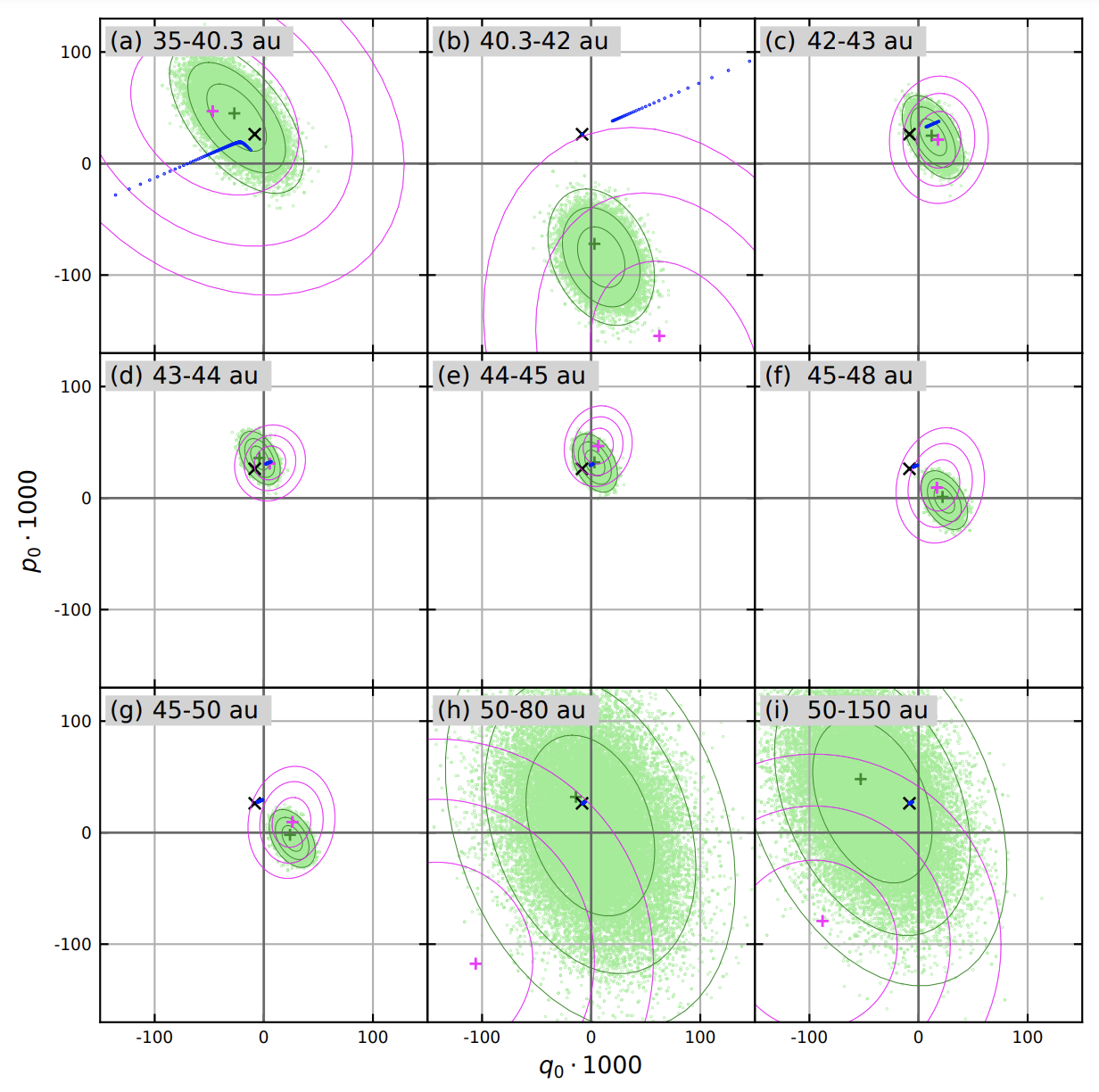}
	\caption{Kuiper Belt mean plane and its confidence ellipses in the $(q,p)$ plane, by semimajor axis bin.  Semimajor axis bins are (a) 35-40.3 au, (b) 40.3-42 au, (c) 42-43 au,  (d) 43-44 au, (e) 44-45 au, (f) 45-48 au, (g) 45-50 au, (h) 50-80 au, and (i) 50-150 au. The best-fit mean plane is the dark green +.  The mean planes of 40,000 Monte-Carlo samples are in light green, and the 68.2\%, 95.4\%, and 99.7\% covariance ellipses for them are in dark green. The J2000 ecliptic/equinox pole is at the origin.  The invariable plane is the black x. The best-fit mean plane from \textsc{vm17} is the magenta +. The theoretical prediction (from linear secular theory) for the Laplace surface as a function of semimajor axis is plotted in blue. More details are given in the main text. The 68.2\%, 95.4\%, and 99.7\% covariance ellipses from \textsc{vm17} are plotted in magenta. Each subplot has the same scale, showing that increased sample size does reduce mean plane uncertainty: in each semimajor axis bin, we have more objects and smaller covariance ellipses than \textsc{vm17}, and the more heavily populated semimajor axis bins on the second row have smaller covariance ellipses than the less heavily populated bins on the first row. The larger covariance ellipses on the third row, above 50 au, reflect the wider model inclination distribution used for those bins.}
	\label{fig:qp_9}
\end{figure}

\begin{figure}[htb!]
	\plotone{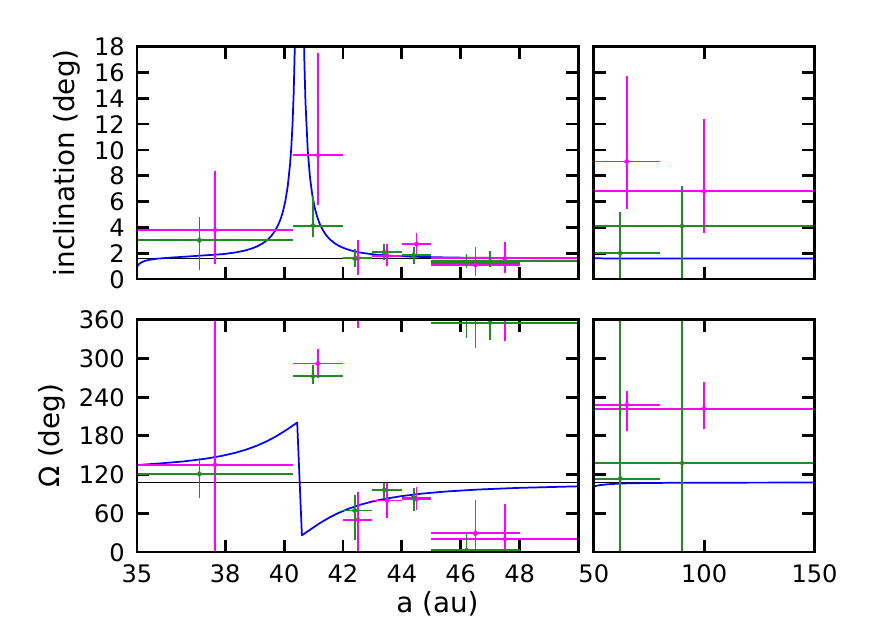}
	\caption{Kuiper Belt mean plane and 68.2\% confidence intervals by semimajor axis bin. The mean planes and confidence intervals from this work are in green, and those from \textsc{vm17} are in magenta.  For the sake of readability, the vertical magenta and green lines have been slightly offset from each other in semimajor axis when they would otherwise overlap. The  theoretical prediction for the Laplace surface as a function of semimajor axis is plotted in blue.  The invariable plane is indicated by the  horizontal black line.}
	\label{fig:iOmega}
\end{figure}

With the non-Resonant objects identified as in Section \ref{sec:sample_selection}, and following the mean plane calculation method described in Section \ref{sec:mean_plane_measurement}, and its confidence interval calculation method described in Section \ref{sec:uncertainty}, the mean planes and confidence intervals we found are shown in graphical form in Figures \ref{fig:qp_2}-\ref{fig:iOmega}, and tabulated in Table \ref{table:vm17method_results}.

In Figure \ref{fig:qp_2}, the left panel shows the results for the classical Kuiper belt (42-48 au, 1242 objects), the center panel shows the results for the entire Kuiper belt (35-150 au, 1812 objects), and the right panel shows the results for the entire Kuiper belt with the classical region excluded (570 objects). The best-fit mean plane is shown as a dark green +, and the 68.2\%, 95.4\%, and 99.7\% confidence ellipses are also indicated in dark green. For comparison, the semimajor axis-dependent Laplace surface (from linear secular theory as developed in \citet{md99}, recomputed with updated planetary parameters for the epoch 2023 February 20) is indicated in blue. For context, we also indicate the location of the invariable plane with a black x. The J2000 ecliptic/equinox pole is located at the origin. We observe that the classical region dominates the results: when it is removed, $i_0$ is nearly unchanged, but $\Omega_0$ shifts by $+67^\circ$, and the dispersion $\sigma_i$ of the synthetic mean planes more than doubles.

In Figure \ref{fig:qp_9}, we plot the results for the mean planes computed at higher resolution in semimajor axis: each panel shows the results for a single semimajor axis bin from Table \ref{table:vm17method_results}. As in Figure \ref{fig:qp_2}, we plot the measured mean plane and its uncertainty distribution, and indicate the invariable plane, and the Laplace surface. In these panels, we also indicate the best-fit mean plane from \textsc{vm17} as a magenta +, with the 68.2\% confidence ellipses from \textsc{vm17} in magenta and with estimated 95.4\% and 99.7\% confidence ellipses scaled from the 68.2\% ellipses.
Note that the Laplace surface appears as a very concentrated set of dots in most panels, but in panels (a) and (b) it is an extended quasi-linear set of dots owing to the warps caused by the $\nu_{17}$ and the $\nu_{18}$ secular resonance, respectively. As $a$ increases from the lower boundary of each  semi-major axis bin to its upper boundary, the Laplace surface traces these paths: in panel (a), the trace begins from the left, at high $i$ and $\Omega\approx 120^\circ$, passes near the origin with $i\approx 1.8^\circ$, and exits to the lower right; in panel (b), the trace begins at the upper right and approaches the invariable plane.

Our results for the mean plane inclination and longitude of node, and their 68.2\% confidence intervals for each semimajor axis bin are tabulated in Table \ref{table:vm17method_results}. These are also plotted in Figure \ref{fig:iOmega} as a function of semimajor axis.  The best-fit mean plane results are  in green, where the horizontal error bars indicate the width of the semimajor axis bin and the vertical error bars indicate the 68.2\% confidence interval.  For comparison, we plot the Laplace surface (varying with semimajor axis) in blue and the invariable plane is indicated by the black horizontal line.  We also indicate the results of \textsc{vm17} in magenta (with a small horizontal offset, for legibility).

\begin{deluxetable}{lDDDDDDDD}[htb] \label{table:vm17method_results}
	\decimals
	\tablecaption{Measured Kuiper Belt Mean Plane}
	\tablehead{\colhead{Semimajor axis bin, au}  & \twocolhead{Count} & \twocolhead{Count (\textsc{vm17})} & \twocolhead{$i_0$ (deg)} &  \twocolhead{$\Omega_0$ (deg)} &  \twocolhead{$\sigma_i$ (deg)}}
	\startdata
	35-40.3           &  92  &  43  & 3.0_{-2.3}^{+1.8} & 121_{-37}^{+25}   & 1.59 \\
	40.3-42           & 164  &  82  & 4.1_{-0.9}^{+2.3} & 272_{-12}^{+17}   & 1.40 \\
	42-43             & 207  & 100  & 1.6_{-0.7}^{+0.8} & 64_{-46}^{+24}    & 0.81 \\
	43-44             & 360  & 186  & 2.1_{-0.6}^{+0.6} & 96_{-15}^{+11}    & 0.54 \\
	44-45             & 273  & 141  & 1.8_{-0.7}^{+0.6} & 85_{-22}^{+15}    & 0.59 \\
	45-48             & 402  & 194  & 1.3_{-0.4}^{+0.7} & 3_{-31}^{+25}     & 0.60 \\
	45-50             & 440  & 217  & 1.4_{-0.4}^{+0.8} & 355_{-26}^{+22}   & 0.59 \\
	50-80             & 221  & 125  & 2.0_{-2.0}^{+3.2} & 114_{-114}^{+246} & 4.02 \\
	50-150            & 276  & 162  & 4.1_{-4.1}^{+3.1} & 138_{-138}^{+222} & 3.66 \\
	Total (35-150)    & 1812 & 931  & 1.6_{-0.4}^{+0.2} & 82_{-15}^{+6}     & 0.28 \\
    Classical (42-48) & 1242 & 617  & 1.6_{-0.4}^{+0.3} & 82_{-14}^{+8}     & 0.31 \\
    Total-Classical   &  570 & 314  & 1.7_{-1.0}^{+0.4} & 149_{-15}^{+39}   & 0.65 \\
	\enddata
	\tablecomments{The best-fit and 68.2\% confidence intervals of the measured mean plane's ecliptic inclination, $i_0$, and longitude of ascending node, $\Omega_0$, and the uncertainty of the pole position, $\sigma_i$.}
\end{deluxetable}

\begin{deluxetable}{lllllll}[htb] \label{table:rejections}
    \tablecaption{Statistical significance of the measured mean plane compared to the Invariable Plane (IP) and Laplace surface (LS)}
    \tablehead{\colhead{Semimajor axis bin, au} & \colhead{IP, 68.2\%} & \colhead{IP, 95.4\%} & \colhead{IP, 99.7\%} & \colhead{LS, 68.2\%} & \colhead{LS, 95.4\%} & \colhead{LS, 99.7\%}}
    \startdata
    35-40.3           & \checkmark & \checkmark & \checkmark & \checkmark & \checkmark & \checkmark \\
    40.3-42           & ...        & ...        & ...        & ...        & ...        & ...        \\
    42-43             & ...        & ...        & \checkmark & \checkmark & \checkmark & \checkmark \\
    43-44             & ...        & \checkmark & \checkmark & \checkmark & \checkmark & \checkmark \\
    44-45             & ...        & \checkmark & \checkmark & \checkmark & \checkmark & \checkmark \\
    45-48             & ...        & ...        & ...        & ...        & ...        & ...        \\
    45-50             & ...        & ...        & ...        & ...        & ...        & ...        \\
    50-80             & \checkmark & \checkmark & \checkmark & \checkmark & \checkmark & \checkmark \\
    50-150            & \checkmark & \checkmark & \checkmark & \checkmark & \checkmark & \checkmark \\
    Total (35-150)    & ...        & ...        & ...        & \checkmark & \checkmark & \checkmark \\
    Classical (42-48) & ...        & ...        & ...        & \checkmark & \checkmark & \checkmark \\
    Total-Classical   & ...        & ...        & ...        & \checkmark & \checkmark & \checkmark \\
    \enddata
    \tablecomments{Acceptance/rejection decisions for invariable plane and Laplace surface compared to the measured mean plane in each semimajor axis bin. Checks are accepted, ellipses are rejected. A check indicates that the 68.2\% (95.4\%, 99.7\%) confidence ellipse of the measured mean plane pole for the semimajor axis bin contains the pole of the invariable plane, or it contains the pole of the Laplace surface for any semimajor axis within the boundaries of the bin. An ellipsis indicates the contrary.}
\end{deluxetable}

Is the measured mean plane consistent with the Laplace surface or with the invariable plane?
For the classical Kuiper Belt from 42-48 au, we find the mean plane of 1242 objects as $i_0=(1.6^\circ)_{-0.4}^{+0.3}$, $\Omega_0=(82^\circ)_{-14}^{+8}$. For the entire belt from 35-150 au, we find the mean plane of 1812 objects as $i_0=(1.6^\circ)_{-0.4}^{+0.2}$, $\Omega_0=(82^\circ)_{-14}^{+8}$. When we exclude
the classical region, we find the mean plane of 570 objects as $i_0=(1.7^\circ)_{-1.0}^{+0.4}$, $\Omega_0=(149^\circ)_{-15}^{+39}$. In all three cases, this is close to the invariable plane (respectively within $0.7^\circ$, $0.6^\circ$, and $1.2^\circ$), but distinguishable from it at greater than 99.7\% confidence, as evident in Figure \ref{fig:qp_2}.

We do not comment on any distinction between the measured mean plane and the Laplace surface for these broad ranges, as the changing location of the Laplace surface renders the question ill-formed, but we can answer this question for the smaller semimajor axis bins. In Table \ref{table:rejections}, we tabulate simple acceptance/rejection decisions with respect to the invariable plane and the Laplace surface for each semimajor axis bin.
At the 99.7\% confidence level, we can only reject the invariable plane or the Laplace surface for the bins 40.3-42, 45-48, and 45-50 au. In each of those bins, we reject both planes. Elsewhere, both planes are consistent with the sample population.

The discrepancy between the theoretically expected Laplace surface and the measured mean plane in the 40.3-42 au bin may arise from inaccuracies of the linear theory for the theoretical estimate of the strong warp in proximity to the $\nu_8$ secular resonance. In the case of the 45-48 au and 45-50 au bins, we note that both these bins are in proximity to Neptune's 2:1 mean motion resonance but the theoretical Laplace surface does not account for the effects of mean motion resonances.

Because our methodology closely follows that of \textsc{vm17} and our main motivation was to test the reproducibility of their results with the updated sample of KBOs, we comment in some detail on the comparisons between those results and ours. In general, the results reported here have smaller measurement uncertainties of the mean pole positions than those in \textsc{vm17}, due to the larger sample sizes now available.
\begin{itemize}
\item In the 35-40.3 au bin (Figure \ref{fig:qp_9}a), almost our entire 99.7\% confidence ellipse falls within their 68.2\% confidence ellipse. While our 68.2\% confidence ellipse is small enough to uniquely identify $\Omega_0$ at that level of confidence, our 99.7\% ellipse surrounds the origin so we cannot report a unique $\Omega_0$ at the higher confidence level.
\item In the 40.3-42 au bin (Figure \ref{fig:qp_9}b), our 68.2\% confidence ellipse falls mostly outside theirs and is closer to the origin, resulting in a similar 68.2\% confidence interval for $\Omega_0$ but a distinctly different (and lower) estimate and interval for $i_0$. Our results are closer to the Laplace surface than theirs, but still differ from the Laplace and invariable planes by more than 99.7\% confidence. With future increases in the size of KBO samples, splitting this bin into two or more narrower semimajor axis ranges could enable higher precision comparisons of the measured and predicted Laplace surface within the constraints of the existing theory. Without splitting the bin, the $\nu_{18}$ secular resonance warps the Laplace surface so strongly with semimajor axis that a second-order theory is necessary for accurate predictions.
\item In the 42-43, 43-44, and 44-45 au bins (Figure \ref{fig:qp_9}c, d, e), our measured mean planes fall within their 68.2\% covariance ellipses, and we produce overlapping, though smaller, 68.2\% confidence intervals.
\item In the 45-48 and 45-50 au bins (Figure \ref{fig:qp_9}f, g), our measured mean planes again fall within their 68.2\% covariance ellipses, but our 95.4\% and 99.7\% covariance ellipses are small enough that, unlike \textsc{vm17}, we reject the Laplace surface and invariable plane at all three confidence levels. These bins are close to Neptune's 2:1 outer mean motion resonance.
\item Most strikingly, we do not detect the strong warp \textsc{vm17} reported for the distant Kuiper Belt near $a>50$ au (Figure \ref{fig:qp_9}h, i). In the 50-80 and 50-150 au bins, the Laplace surface and invariable plane are both within $2.9^\circ$ of our measured mean planes, and fall within our 68.2\% covariance ellipses, while for \textsc{vm17} they are more than $6^\circ$  away from the measured mean planes and fall near the 99.7\% covariance ellipses.

We cannot currently rule out the invariable plane, the Laplace surface, or even the ecliptic plane as the mean plane of the Kuiper belt in this region at even the 68.2\% confidence level. We note that our sample size in this region is 272, which is 70\% larger than \textsc{vm17}'s sample. Although the best-estimate of the mean plane of the distant Kuiper belt differs by more than $2\sigma$ in the two studies, there is significant overlap of the $3\sigma$ confidence ellipses of our and their measurements. Larger sample sizes in the future would help to better understand this region by reducing the measurement uncertainties. If we assume, as a rule of thumb, that the size of the uncertainty ellipse varies with the sample size as $\sim N^{-1/2}$, then reducing the measurement uncertainty by a factor of 2--3 will require a sample size 4--9 times as large. This is likely to be achieved over the next decade as the Vera C. Rubin Observatory carries out the Legacy Survey of Space and Time \citep[LSST, ][]{ivezic2019}.

\end{itemize}

We briefly comment on comparisons of our results with those of other previous studies.
\begin{itemize}
\item \cite{cc08} computed the Kuiper Belt mean plane and its uncertainty for 10 objects with $38.09 < a < 39.10$ au, $e<0.1$, and ecliptic inclination $i<10^\circ$, and for 80 objects with $42.49 < a < 43.50$ au and the same eccentricity and inclination cutoffs. They did not rule out either the Laplace surface or the invariable plane as the true Kuiper Belt plane for those two specific semimajor axis ranges with greater than $3\sigma$ (99.7\%) confidence.

When we impose the same eccentricity and inclination restrictions on our sample, we respectively obtain 47 and 347 objects in these semimajor axis ranges. When we compute the mean plane and its uncertainty for these samples, we do not rule out the Laplace surface at any confidence level for either sample, and we do not rule out the invariable plane for either sample at a confidence level above 68.2\%. Removing the eccentricity and semimajor axis cutoffs adds no objects to either semimajor axis range, and does not change the outcome.

\item \cite{ossos14} computed the plane of the cold classical Kuiper Belt, with uncertainties, for 107 objects in the 42.4-43.8 au bin, 82 objects in the 43.8-44.4 au bin, and 67 objects in the 44.4-47 au bin, where a cold object was defined as having an inclination of $4^\circ$ or less relative to the Laplace surface at the center of the semimajor axis bin (i.e. 43.1 au for the 42.4-43.8 au bin). They computed the Kuiper belt plane and its uncertainties for an unspecified number of hot classical objects between 42.4-47 au, where a hot object has an inclination of $9^\circ$ or more relative to the Laplace surface at 44.7 au.
They repeated the computations, without inclination restrictions, for 57 objects in the 50-80 au bin and for 83 objects in the 48-150 au bin.  In the 42.4-43.8 au and 43.8-44.4 au bins of cold objects, they rejected the invariable plane but not the Laplace surface at 99\% confidence. In the 44.4-47 au bin of cold objects, the 42.4-47 au bin of hot objects and the higher--$a$ bins ($a > 47$~au), they rejected neither the invariable plane nor the Laplace surface at 99\% confidence.

When we impose the same semimajor axis and inclination restrictions on our sample, we find 450 cold objects in the 42.4-43.8 au range, 284 cold objects in the 43.8-44.4 au range, and 444 cold objects in the 44.4-47 au range. We find 1178 hot objects in the 42.4-47 au range, and 314 total objects in the 48-150 au range. In each case, our sample sizes are larger than \cite{ossos14}'s by factors of three or more. In the 42.4--43.8 au bin of cold objects, we reject the invariable plane at 95.4\% confidence but do not reject the Laplace surface at any confidence level. In the 43.8-44.4 au bin of cold objects, we reject the invariable plane at 68.2\% confidence but do not reject the Laplace surface at any confidence level.
These results are similar to those shown for nearby semimajor axis bins in Figure \ref{fig:qp_9}.
In the 44.4-47 au bin of cold objects, we reject neither the invariable plane nor the Laplace surface at any confidence level. In the 42.4-47 au bin of hot objects, we reject both the invariable plane and the Laplace surface (for 44.7 au, the center of the bin) at well above 99.7\% confidence. As previously noted, for the higher-$a$ bins we reject neither the invariable plane nor the Laplace surface at 95.4\% confidence.

Our results for the cold population agree with those of \citet{ossos14}, but we reject at high confidence both the invariable plane and the Laplace surface for the hot objects in the 42.4-47 au range where they rejected neither. Our simulated mean planes in that bin are tightly clustered around our observed mean plane, while theirs were tightly clustered around the invariable plane. We suspect that this discrepancy is explainable by our much larger sample size; perhaps the mean plane of their much smaller sample was located much closer to the invariable plane.

\item \cite{bp04} rejected the invariable plane at the $3\sigma$ (99.7\%) confidence level for the Kuiper Belt as a whole. We also reject it for the entire belt between 35-150 au.

\item \cite{e05} rejected the Laplace surface at the $1\sigma$ (68.2\%) confidence level for 85 ``Classical" objects in the range $37.9<a<47.0$ au. In their definition, a Classical object is a non-Resonant object with $e\leq 0.2$ and a Tisserand parameter relative to Neptune of $T_N>3$.

When we impose the same eccentricity and Tisserand restrictions on this semimajor axis range, we obtain 1507 objects. When we compute the mean plane and its uncertainty for these 1451 objects, we reject the invariable plane at 99.7\% confidence but do not reject the Laplace surface at any confidence level. Removing the eccentricity and Tisserand cutoffs adds no objects to the semimajor axis range, and does not change the outcome.
\end{itemize}

We close with the observation that future increases in the  sample sizes of outer solar system minor planets will enable higher confidence results in measurements of the Kuiper Belt mean plane, and should be examined for their potential to detect the effects of unseen distant planets or other unmodeled perturbations.

%\begin{acknowledgments}
\vspace{12pt}

We thank Aaron Rosengren, James Keane, and especially Kathryn Volk for correspondence, comments, and sample code. We acknowledge funding from the NSF (grant AST-1824869) and from the JPL SURP and SIP student support programs.  We also thank the anonymous reviewer and the data editor and statistics editor for their comments, questions, and advice.

%\end{acknowledgments}

\vspace{5mm}

\software{astroquery \citep{2019AJ....157...98G},
	      rebound \citep{rebound},
	      scikit-learn \citep{scikit-learn},
	      shapely \citep{shapely2007},
            KBO classifier \citep{sv20}
          }

\bibliography{references}
\bibliographystyle{aasjournal}

\end{document}